%% file: Cygx1.tex
\documentclass{aastex631}

\shorttitle{An X-ray Spectro-Polarimetric Campaign of Cyg X-1's Soft State}
\shortauthors{Steiner et al.}

\usepackage{amsmath}

\usepackage{xcolor}

\usepackage{graphicx}
\usepackage[space]{grffile}
\usepackage{latexsym}
\usepackage{textcomp}
\usepackage{longtable}
\usepackage{multirow,booktabs}
\usepackage{amsfonts,amsmath,amssymb}
\usepackage{natbib}
\usepackage{url}
\usepackage{hyperref}

\newcommand{\msun}{M_\odot}

\newcommand{\spin}{a_*}

\newcommand{\nh}{N_{\rm H}}

\newcommand{\cyg}{{Cyg~X-1}}
\newcommand{\Cyg}{{Cyg~X-1}}

\newcommand{\nicer}{{NICER}}
\newcommand{\NICER}{{NICER}}
\newcommand{\INTEGRAL}{{INTEGRAL}}
\newcommand{\MAXI}{{MAXI}}
\newcommand{\ixpe}{{IXPE}}
\newcommand{\IXPE}{{IXPE}}

\newcommand{\NuSTAR}{{NuSTAR}}
\newcommand{\NUSTAR}{{NuSTAR}}
\newcommand{\Swift}{{Swift}}

\newcommand{\AstroSAT}{{AstroSat}}

\begin{document}

\title{An IXPE-Led X-ray 
Spectro-Polarimetric Campaign on the Soft State of Cygnus X-1: X-ray Polarimetric Evidence for Strong Gravitational Lensing}

\author[0000-0002-5872-6061]{James F. Steiner}
\affiliation{Center for Astrophysics \textbar\ Harvard \& Smithsonian, 60 Garden St, Cambridge, MA 02138, USA}

\correspondingauthor{James F. Steiner}
\email{jsteiner@cfa.harvard.edu}

\author[0000-0002-9633-9193]{Edward Nathan}
\affiliation{California Institute of Technology, Pasadena, CA 91125, USA}

\author[0000-0002-9705-7948]{Kun Hu}
\affiliation{Physics Department, McDonnell Center for the Space Sciences, and Center for Quantum Leaps, Washington University in St. Louis, St. Louis, MO 63130, USA}

\author[0000-0002-1084-6507]{Henric Krawczynski}
\affiliation{Physics Department, McDonnell Center for the Space Sciences, and Center for Quantum Leaps, Washington University in St. Louis, St. Louis, MO 63130, USA}

\author[0000-0003-0079-1239]{Michal Dov\v{c}iak}
\affiliation{Astronomical Institute of the Czech Academy of Sciences, Boční II 1401/1, 14100 Praha 4, Czech Republic}

\author[0000-0002-5767-7253]{Alexandra Veledina}
\affiliation{Department of Physics and Astronomy, 20014 University of Turku, Finland}
\affiliation{Nordita, KTH Royal Institute of Technology and Stockholm University, Hannes Alfvéns väg 12, SE-10691 Stockholm, Sweden}

\author[0000-0003-3331-3794]{Fabio Muleri}
\affiliation{INAF Istituto di Astrofisica e Planetologia Spaziali, Via del Fosso del Cavaliere 100, 00133 Roma, Italy}

\author[0000-0003-2931-0742]{Jiri Svoboda}
\affiliation{Astronomical Institute of the Czech Academy of Sciences, Boční II 1401/1, 14100 Praha 4, Czech Republic}

\author[0000-0003-0168-9906]{Kevin Alabarta}
\affiliation{Center for Astrophysics and Space Science (CASS), New York University Abu Dhabi, PO Box 129188, Abu Dhabi, UAE}

\author[0009-0003-8610-853X]{Maxime Parra}
\affiliation{Universit\'{e} Grenoble Alpes, CNRS, IPAG, 38000 Grenoble, France}
\affiliation{Dipartimento di Matematica e Fisica, Universit\`a degli Studi Roma Tre, Via della Vasca Navale 84, 00146 Roma, Italy}

% ...


\begin{abstract}
    We present the first X-ray spectropolarimetric results for Cygnus X-1 in its soft state from a campaign of five \IXPE\ observations conducted during 2023 May--June.  Companion multiwavelength data during the campaign are likewise shown.   The 2--8 keV X-rays exhibit a net polarization degree PD=$1.99\%\pm0.13\%$ (68\% confidence). 
     The polarization signal is found to increase with energy across \IXPE's 2--8 keV bandpass. The polarized X-rays exhibit an energy-independent polarization angle of PA=$-25\fdg7 \pm 1\fdg8$
    East of North (68\% confidence).  This is consistent with being aligned to \cyg's AU-scale compact radio jet and its pc-scale radio lobes. 
     In comparison to earlier hard-state observations, the soft state exhibits a factor of 2 lower polarization degree, but a similar trend with energy and a similar (also energy-independent) position angle.
    When scaling by the natural unit of the disk temperature, we find the appearance of a consistent trendline in the polarization degree between soft and hard states. Our favored polarimetric model indicates \Cyg's spin is likely high ($\spin \gtrsim 0.96$).
    The substantial X-ray polarization in Cyg X-1's soft state is most readily explained as resulting from a large portion of  X-rays emitted from the disk returning and reflecting off  the disk surface, generating a high polarization degree and a polarization direction parallel to the black hole spin axis and radio jet.  In \IXPE's bandpass, the polarization signal is dominated by the {\em returning} reflection emission.   This constitutes polarimetric evidence for strong gravitational lensing of X-rays close to the black hole.   
\end{abstract}

\section{Introduction}\label{sec:intro}

In a recent pioneering study of the hard state of Cygnus~X-1 (hereafter, \cyg), the Imaging X-ray Polarimetry Explorer \citep[\IXPE;][]{Weisskopf22} produced the first unambiguous soft X-ray polarimetric detection of a black hole (BH), and moreover measured a significant increase in the degree of polarization across \IXPE's 2--8 keV bandpass \citep{Henric_ixpe_cygx1}.  An accompanying set of \NICER\ and \NuSTAR\ observations spanning the campaign proved key to deciphering the \IXPE\ results by identifying the source of polarized emission (the corona, or possibly a hot flow) and thereby enabling a spectro-polarimetric constraint on the geometry of the inner accretion flow (i.e., the inner disk and its enshrouding hot-electron corona).  These first-of-their-kind measurements included several surprising results, foremost the unexpectedly {\em strong} polarization at 4\%.  These results were able to unambiguously rule out (for the hard state) a popular ``lamppost'' model for the corona's geometry in \Cyg.  Here, we report a cousin campaign again led by \IXPE\ to explore the X-ray polarimetric signature of \cyg's soft state.

\Cyg\ is the brightest persistent BH source in the Galaxy at $\sim$ 0.2--2~Crab (2--20~keV), and is the first X-ray binary widely accepted to harbor an accreting BH \citep{Bolton_1972,Webster_Murdin_1972}. 
Its X-ray emission is fueled by the accretion of powerful winds from its O-supergiant companion \citep{Orosz_2011}.  A recent parallax study by \citet{JMJ_Cygx1}  yielded a precise distance of $D=2.2\pm0.2$~kpc, and a refined BH mass $M=21.2\pm2.2~\msun$ with a companion mass of $41\pm7~\msun$.  This marks \Cyg\ as the most massive BH among currently known X-ray binaries.

From \Cyg's proper motion in the Galaxy, \citet{JMJ_Cygx1} strongly constrain \cyg's natal kick,  $v<10-20$~km s$^{-1}$ (see also \citealt{Mirabel_2003}).  This indicates that any misalignment of the spin and binary orbital angular momentum of the nascent \cyg\ should be slight, at most $\phi \lesssim 10\degr$ when formed.  This result is significant in light of a precise orbital inclination measurement via ellipsoidal light-curve modeling \citep{Orosz_2011}, $i_{\rm orb}\approx 27\degr\pm1\degr$.

Across decades of X-ray monitoring, \Cyg\ has been found to range by an order of magnitude in its X-ray flux, typically accreting at a few percent of its Eddington limit.  Such a range is, however, remarkably constant in comparison to the $>6$ orders of magnitude traversed in the outburst of a typical BH transient.  At the same time, unlike the other wind-fed BHs, \Cyg's low mass accretion rate causes it to range along the lower track of BH hardness-intensity, producing intermittent state transitions. 

In fact, \Cyg's state changes were prototypical for establishing a hard/soft state dichotomy \citep{Tananbaum_1972}.  While this behavior served as archetype for establishing the hard and soft spectral states ubiquitous among X-ray binaries  \citep{Oda_1977}, its soft state never reaches the extremity of the long-lived ``thermal-dominant'' state common among BH X-ray transients, in which the coronal X-ray contribution is minimal.  Instead, \Cyg's soft state is associated with the canonical ``steep power law'' (SPL) or sometimes ``soft-intermediate'' state \citep{RM06, Fender_2004}.

Broadly, hard states are dominated by emission Comptonized in a hot corona, and exhibit radio jets, whereas soft states are dominated by thermal-disk emission and present weak or no  radio-jet activity \citep{Fender_2004}. \Cyg\ regularly presents a compact radio jet in its soft state \citep{Zdziarski_2020}.  Compared to hard states, the soft state disk temperature is generally higher while the coronal emission is both weaker and spectrally steeper.  Wide-ranging evidence on spectral and timing grounds implies that state transitions are associated with significant changes in the geometry of the innermost accretion flow (e.g, \citealt{Ichimaru_1977, DeMarco_2015,Jingyi_2022, Mendez_2022}).  At the same time, the nature of such structural changes as well as the location and shape of the corona are contentious, possibly involving the truncation of the inner accretion disk at radii much larger than the innermost stable circular orbit (ISCO; e.g. \citealt{Basak_Zdziarski, Garcia_GX}) and with candidate coronal geometries ranging from sphere to slab to lamppost (e.g., \citealt{Dove_1997, compps, Dauser_2016}). 

\Cyg's hard and soft states commonly persist for weeks to years at a time.  Despite the slow pace of inter-state evolution, both its hard and soft states exhibit substantial intra-state secular variations in intensity and hardness over timescales from hours to days, and its power-density spectrum shows a pronounced broad feature near $\sim 1$~Hz (e.g., \citealt{Grinberg_2014}).  
In the soft state, \Cyg's disk is expected to reach the ISCO. The disk produces a quasi-black body spectrum which dominates $\lesssim3$~keV with higher energies indicating the presence of a corona with hybrid thermal/non-thermal electrons \citep{Poutanen1998,Gierlinski_1999}.  The disk-dominated spectrum can be modeled via continuum fitting to determine the radius of the ISCO which is linked to BH spin \citep{Zhang_1997, Gou_2014}. 
The ISCO radius is also traced by the relativistic broadening of ``reflection'' features.  These were first discovered in \Cyg\ \citep{Barr_1985} and arise from X-ray irradiation reprocessing in the accretion disk's surface \citep{Fabian_1989}.  
Typically, this X-ray source is the corona, but could in some instances be \emph{returning radiation} from the disk illuminating itself  \citep{Cunningham_1976, Connors_2020, KrawczynskiBeheshtipour+22}.

  Numerous X-ray spectroscopic studies have explored the question of \Cyg's dimensionless spin parameter ($\spin \equiv \frac{cJ}{GM^2}$).  Continuum-fitting studies focused on the thermal disk emission in soft states \citep{Zhang_1997, McClintock_2006}, consistently find a near-maximal value ($\spin \gtrsim 0.98$; \citealt{Gou_2011, Gou_2014, Zhao_CygX1_2021}). This conclusion has been supported by numerous reflection studies as well, including analyses of both hard and soft states  (e.g., \citealt{Fabian_2012, Tomsick_2014, Walton_2016, Basak_2017, Tomsick_2018}),  but see  \citet{Zdziarski_2024}. Some reflection fits suggest a possible $\sim 10\degr$ misalignment with the binary orbital plane, an interpretation also favored in our hard-state polarimetric study \citep{Henric_ixpe_cygx1}.

Before \IXPE's measurements, soft X-ray polarization in \cyg\ was {\em tentatively} detected ($<99$\% confidence) by OSO-8 with polarization degree (PD) $\sim$2\%--5\%  \citep{Weisskopf_1977,Long_1980}.
At energies above $\gtrsim$200~keV, polarization for \cyg's hard state has been studied using \INTEGRAL, with ISGRI and SPI instruments independently \citep{Laurent_2011, Jourdain_2012, Rodriguez_2015}, and more recently by the \AstroSAT-CZTI in the 100--380~keV band \citep{Chattopadhyay2024}. At high energies, \cyg\ is found to be strongly polarized, and the PD rising with energy. It is most prominent in the so-called ``hard tail'' above $\gtrsim$400 keV (i.e.,  excess emission above the Comptonization cutoff), where  the PD exceeds 60\%.  The PoGO+ balloon-borne polarimeter constrained \cyg's PD$<8.6\%$ in the 19--181\,keV range for its hard state \citep{Chauvin_2018}. Polarization in the soft state at high energies could not be constrained with current \INTEGRAL\ data due to low signal. Although a hard tail has been detected in the soft state \citep{McConnell_2002}, constraints on its properties are model dependent (\citealt{Cangemi_2021}, see also \citealt{Chattopadhyay2024}).

This Letter is organized as follows:  In Section~\ref{sec:data} we detail our observational campaign and present the data.  We show fits to these data and associated spectro-polarimetric results in Section~\ref{sec:results}.  We offer a discussion of these results in Section~\ref{sec:discuss}, and concluding remarks in Section~\ref{sec:conc}.

\section{Data}
\label{sec:data}

In April 2023, \cyg\ transitioned from a long-lived  hard state to the soft state, initiating a corresponding \IXPE\ soft-state monitoring campaign.  In order to prevent data-recorder overflow, the observation was broken into 5 epochs spanning 2023 May 02 through June 20.   A preliminary look at the polarimetric results was posted in an Astronomer's Telegram (AT) \citep{CygX1_ATEL}, motivating a rich multi-wavelength campaign to supplement \IXPE's later epochs including complementary X-ray spectral coverage during Epochs 3--5 with \NuSTAR, \NICER, \INTEGRAL, \AstroSAT, and \Swift, as summarized in Table~\ref{tab:observations}.  
\IXPE\ polarimetric results from Epochs 1-3 were published by \cite{Jana_2024}; their presentation of the polarimetric measurements are aligned with those of the AT and what we present here, although our conclusions differ from the ones of \cite{Jana_2024}.

We focus our spectroscopic analysis on Epochs 3–5 for which we have complementary broadband X-ray coverage.  Low-energy data are particularly important to constrain \cyg's thermal disk emission which exhibits peak temperatures $\lesssim 0.5$\,keV.  The broadband spectral data in combination with \IXPE's spectro-polarimetric information allow us to identify and constrain the different emission components. The net polarization of the soft/thermal state in the \IXPE\ band depends on the polarization of the thermal disk emission, which is thought to exhibit a PD minimum and a large-amplitude (90\degr) swing of the PA just above the thermal peak (e.g., \citealt{Schnittman_2009}), and is also very sensitive to Compton-scattering in the corona \citep{Schnittman_2010}.

Broadband X-ray fits require coverage by  missions other than \IXPE\  and are crucial to disentangling the components of \cyg's emission, particularly given that \IXPE\ has coarse energy resolution ($\sim$20\%), and is sensitive over a limited range of 2--8~keV.  Compared to this, \Cyg's disk emission is cool and out of band, with temperatures $\lesssim 0.5$~keV.\footnote{The  average disk fractions (by flux) for these soft states are: $\approx 60\%$ from 2--4~keV,  $\lesssim 10\%$ from 4--6 keV, and $<1\%$ above 6~keV.} Accordingly, in Section~\ref{sec:results} our spectroscopic analysis focuses on Epochs~3--5.  

\begin{figure}[t]
  \includegraphics[clip=true, trim={120 88 90 97}, width=0.49\linewidth]{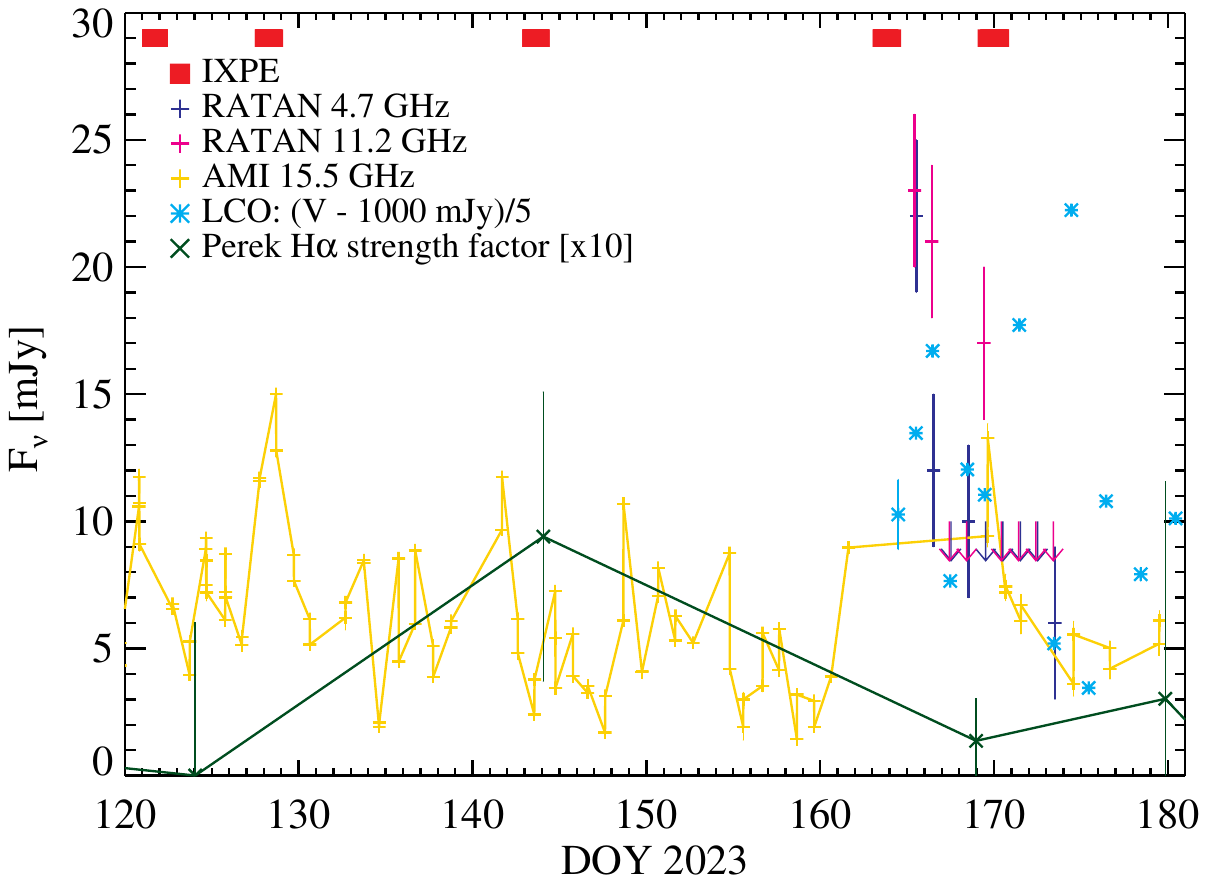}
  \includegraphics[clip=true, trim={120 90 90 102}, width=0.49\linewidth]{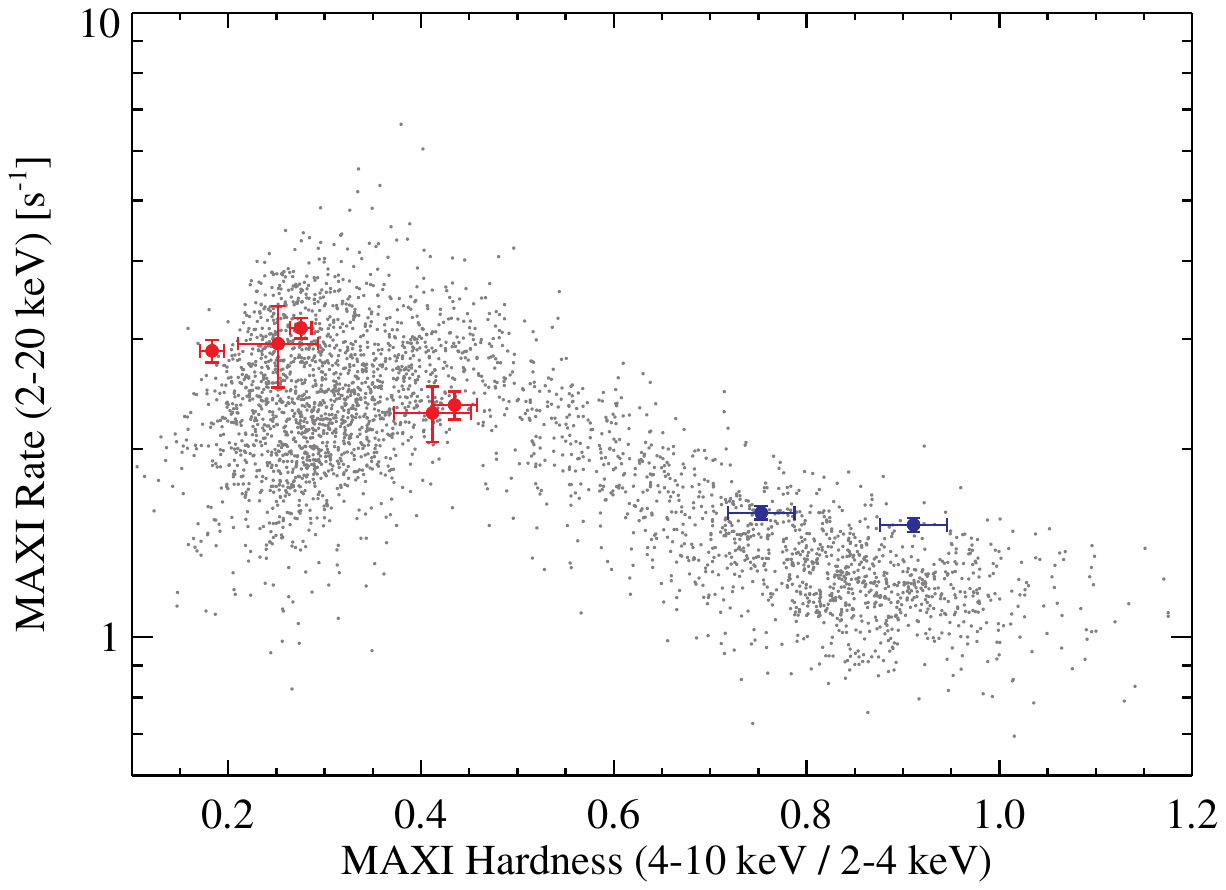}
\caption{
     ({\em Left}) \cyg\ light curve in optical and radio over the span of the \ixpe\ campaign (red bars).  Radio data from AMI and RATAN reveal ongoing  weak jet activity, with a potential flare near Epoch 4.  V-band optical data from LCO shows structured variation at low amplitude ($\lesssim 10\%$), with emission dominated by the companion star.   The same substructure also appears in B, r$^\prime$, and i$^\prime$ bands (not shown here).  H$\alpha$ line-strength factors from the Perek Telescope are shown in green. The larger time baseline of this monitoring reveals that high line strength factor is associated with a decrease in X-ray and increase in radio flux (increasing H$\alpha$ in the companion's stellar wind). 
Light curves are in mJy unless otherwise indicated.  
 ({\em Right}) MAXI hardness-intensity diagram of \cyg, with the 5 \IXPE\ soft-state epochs marked in red and the two hard states in blue. }         
     \label{fig:mw}
\end{figure}

The \ixpe\ campaign was also supplemented with multiwavelength monitoring  carried out in the radio, with RATAN-600 and AMI, with optical data from Las Cumbres Observatory (LCO) and with the Perek Telescope observing in narrow-band H$\alpha$.  
Figure~\ref{fig:mw} presents  radio and optical measurements of \cyg\ during the \IXPE\ campaign.  The radio jet, while faint, is significantly detected throughout.  The H$\mathrm{\alpha}$ emission originates in the stellar wind and has been separated from emission by circumstellar matter and telluric lines using the disentangling method (see \citealt{Hadrava_1997} and \citealt{Hadrava_2009}).  These line-strengths are an order of magnitude weaker than comparable data from the hard state.

In the right-hand panel, daily \MAXI\ hardness and intensity measurements are shown, illustrating \Cyg's state bimodality, with the \IXPE\ observations overlaid. 
Further details on each data set, including reduction procedures, are given in Appendix~\ref{sec:app:data}.

\begin{deluxetable}{lcccccccc}
\tabletypesize{\footnotesize}
\tablecolumns{9}
\tablewidth{0pc}
\tablecaption{X-ray Observations}
\tablehead{ Epoch & Date (UTC) & Orbital Phase & Mission & Instrument & Exposure Times & \IXPE\ Count Rate & \IXPE\ Hardness  \\
 & & & & & ks & s$^{-1}$ [2--8 keV] & [4--8 keV / 2--4 keV] 
}  \startdata
 1  & 2023 May 02--03  & 0.81--0.90 &  \IXPE\  & GPD & 20.9  & 99  & 0.077 \\     
& \\
2  & 2023 May 09--10  & 0.97--0.09 &  \IXPE\  &  GPD & 31.0  & 124 & 0.074 \\     
    &                  &            &  \Swift\tablenotemark{a}
 & XRT &  1.8    &    &        \\
& \\
 3  & 2023 May 24--25  & 0.71--0.83 &  \IXPE\  &  GPD & 24.8  & 166 & 0.042 \\     
    &                  &            &  \AstroSAT\ & SXT/LAXPC  &  8.6/24.6  &     & \\ 
    &                  &            &   \NICER\    &  XTI &  6.8   &     & \\ 
    &                  &            &  \NuSTAR\    &  FPMA+B & 13.8    &     & \\ 
& \\
4  & 2023 June 13--14 & 0.31--0.43 &  \IXPE\  &  GPD & 28.7  & 146 & 0.057 \\     
    &                  &            & \INTEGRAL\ &  IBIS & 24.7 &     & \\ 
    &                  &            & \NuSTAR\ &   FPMA+B & 9.5   &     & \\ 
&\\
5  & 2023 June 20     & 0.39--0.54 &  \IXPE\  &  GPD & 34.6  & 183 & 0.066 \\     
    &                  &            & \INTEGRAL\ & IBIS & 24.2  &     & \\ 
    &                  &            & \NICER\  &  XTI &  14.5  &     & \\ 
    &                  &            & \NuSTAR\ &  FPMA+B &   10.5  &     & \\ 
\enddata
\tablenotetext{a}{\Swift\ observed \Cyg\ with XRT in windowed-timing mode; however the data were heavily contaminated by photon pile-up and did not yield reliable spectroscopy.  While listed here for completeness, \Swift\ XRT was not included in our analysis.}
\tablecomments{\IXPE\ exposure times have been corrected for detector dead time and count \IXPE\ rates are for \IXPE's default ``NEFF'' weighting.  Orbital phases are given from superior conjunction with the BH using the ephemeris of \citet{Brocksopp_1999}.}  
\label{tab:observations}
\end{deluxetable}

\begin{figure}[t]
  \includegraphics[clip=true, trim={180 32 5 35}, width=0.49\linewidth]{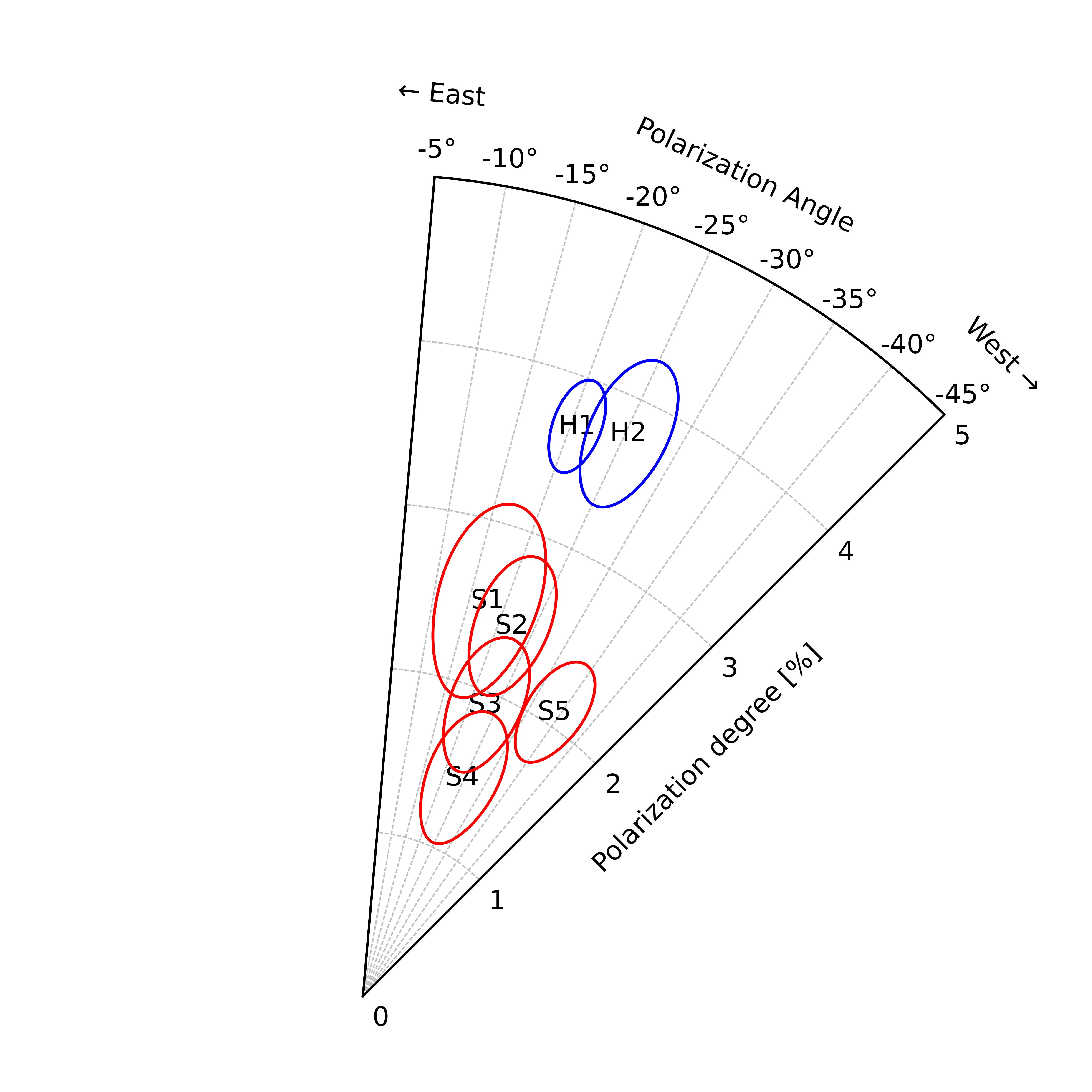}
  \includegraphics[clip=true, trim={180 32 5 35}, width=0.49\linewidth]{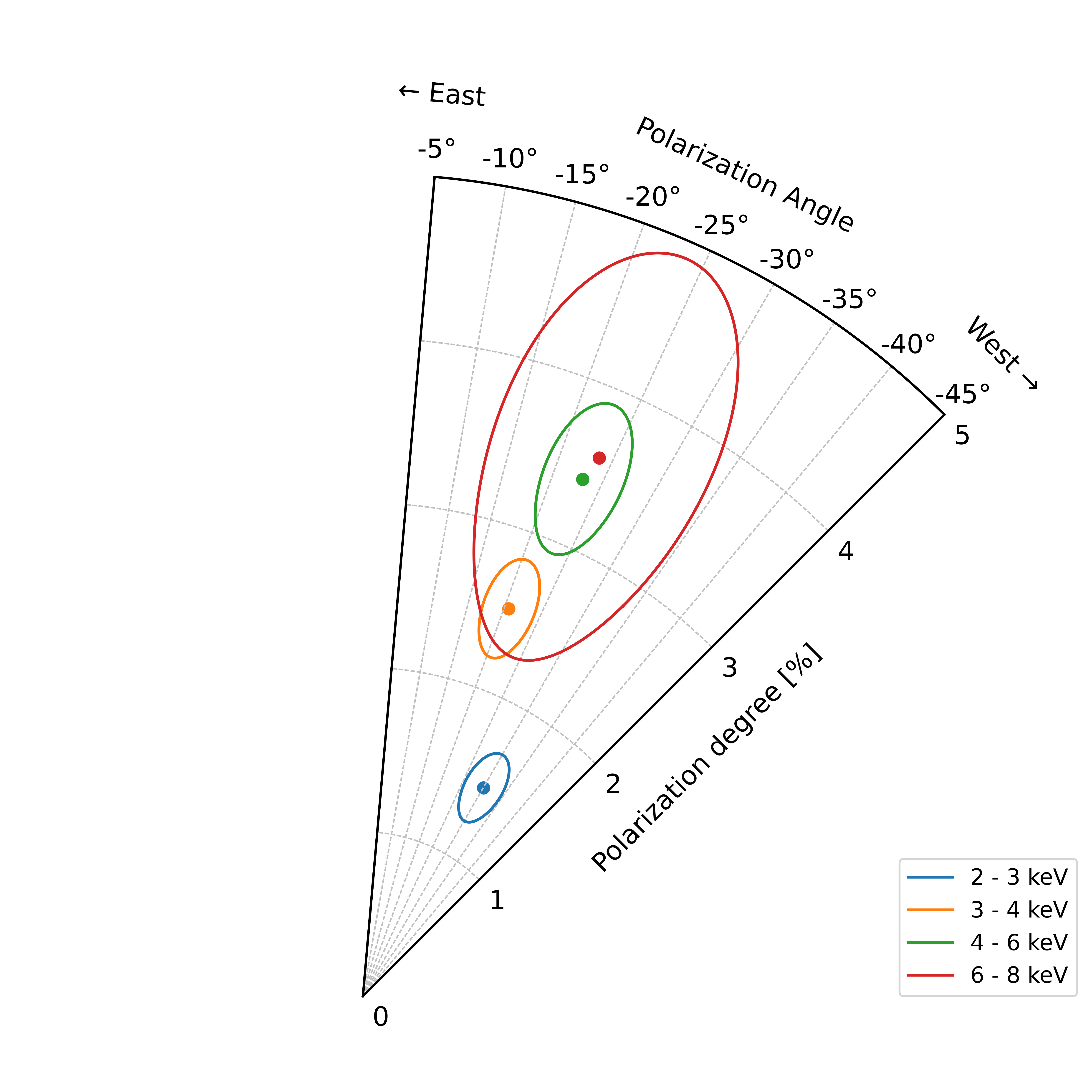}
\caption{({\em Left}) \IXPE\ 2--8 keV PD and PA for each soft state (S1-S5, red) in comparison to previous hard-state observations (H1-S2, blue).  ({\em Right}) Energy-dependent polarization of the soft-state data composite.  Contours show 68\% confidence intervals.
}
\label{fig:polplot}
\end{figure}

\section{Results}
\label{sec:results}

\subsection{Polarimetric Data}
\label{subsec:ixpe}
\IXPE\ observations were processed using the {\sc ixpeobssim} software \citep{ixpeobssim}. For each observation, source events were extracted from an $80\arcsec$ aperture centered on the source. 
Using {\sc ixpeobssim} we obtain the background-subtracted Stokes parameters (see Section~\ref{app:ixpe}), and calculate the PD and the polarization angle (PA) per detector and energy interval of interest.

Figure~\ref{fig:polplot} presents the PD and PA for \cyg\ across the 5 epochs, as determined from Stokes $Q$-and-$U$ data.  The left-hand panel presents polarization of each epoch for the full 2--8~keV range.  The right-hand panel illustrates the energy-dependence of these quantities.  Most critically, while polarized X-rays are detected with high significance, they yield a markedly lower PD compared to the hard state (viz. $2\%$ versus $4\%$). For the soft state data, the net 99\% confidence minimum-detectable polarization is $0.39\%$. The PD increases significantly with energy as was also found in the hard state.  The PA is constant over energy and in time (PA $\approx -26\degr$),  just as seen for the hard state \citep{Henric_ixpe_cygx1}, and aligned to \cyg's  radio jet and radio lobes \citep{Stirling_2001}.

\subsection{Spectral Analysis}
\label{subsec:spectral}

  While coronal emission canonically dominates hard-state X-ray spectra of X-ray binaries, even in \Cyg's soft states, the coronal emission contributes appreciably to the X-ray signal. Because of this, reflection features, including the relativistically-broadened Fe-K$\alpha$ fluorescence line at $\sim 6.5$~keV, and a ``Compton hump'' at $\sim 30$~keV, are correspondingly pronounced in the spectra.  Accordingly, spectral models 
must contain both coronal and reflection components in addition to the thermal-disk.

\subsubsection{Phenomenological Model}
\label{sub:subs:diskbb}
An initial phenomenological spectral analysis of \cyg\ is first pursued using the model: \texttt{tbFeO}$\times ($\texttt{diskbb}+\texttt{smedge}$\times$\texttt{nthcomp}+\texttt{laor}$)$.  Here, \texttt{tbFeO} \citep{tbabs} describes absorption by the insterstellar medium while allowing for nonstandard Fe and O abundances.  The multicolor disk emission is provided by \texttt{diskbb} and the coronal Compton emission by \texttt{nthcomp} \citep{nthcomp, nthcomp2}. Reflection is approximated by the combination of a smeared Fe-edge (\texttt{smedge}; \citealt{smedge})  and the inclusion of a broadened Fe line (\texttt{laor}; \citealt{laor}).
We account for a tear in the thermal shielding of \NuSTAR's FMPA via \NuSTAR's \texttt{nuMLIv1} model \citep{Madsen2020}.

This model has been applied separately for each epoch and although it falls short in capturing the detailed reflection structure, most importantly, it reasonably fits the continuum. 
These empirical fits have $\chi^2$~/~DOF fit statistics of $3484.7 / 2767$, $771.3 / 791$, and $5126.6 / 3836$ when fitting the data in Table~\ref{tab:observations} for Epochs 3, 4, and 5, respectively.
We find that the inner temperature of the \texttt{diskbb} component is $0.506\pm0.003$~keV for Epoch 3, $0.479^{+0.004}_{-0.007}$~keV for Epoch 4, and $0.492^{+0.003}_{-0.005}$~keV for Epoch 5.  Over the full \IXPE\ band, the ratio of thermal (disk) to nonthermal (corona plus reflected) flux is 1.0, 0.82, and 0.58 for Epochs 3, 4, and 5, respectively.


\subsubsection{Fully-relativistic Model}
\label{sub:sub:diskbb}
We next employ a fully-relativistic spectral model in which we replace \texttt{diskbb} with \texttt{kerrbb} \citep{kerrbb}, replace \texttt{nthcomp} with the coronal scattering kernel \texttt{simplcut} \citep{Steiner_simpl, Steiner_2017}, and use \texttt{relxillCP} \citep{relxill,relxill2} to produce reflection emission. Distant reflection from the companion star or a disk rim is included via \texttt{xillverCp}.  In addition, wind absorption is incorporated using \texttt{zxipcf} \citep{zxipcf}.  The wind features are unconstrained without low-energy coverage, and so for Epoch 4, we assume the absorption and wind parameters are the same as in Epoch 5, which was similar in orbital phase.
A spectral-hardening factor in \texttt{kerrbb} describes the ratio of color-to-effective temperature for the thermal disk.  These are decoupled primarily as a result of strong electron scattering in the disk atmosphere. 
 The factor was determined to be $f_{\rm col}=1.55$ using the disk-atmosphere model \texttt{bhspec} \citep{bhspec} for \Cyg's parameters at the temperature and luminosity of our observations, and so we adopt this value throughout. For \texttt{kerrbb}, we fix the BH mass to $21.2\msun$ and distance to $2.22$~kpc \citep{JMJ_Cygx1}, and assume the disk is aligned with the binary inclination of $27\fdg1$ \citep{Orosz_2011}\footnote{The disk rotates clockwise, which then corresponds to $i_{\rm orb}=153\degr$, though in this work we will simply adopt a 0--90$\degr$ range convention referring to the inclination magnitude.}
.  We note that parameter degeneracy within the continuum-fitting disk model prevents a reliable inclination fit for these data (e.g., \citealt{Gou_2011, Steiner_j1550spin}). We apply \texttt{kerrbb}'s returning radiation flag, but not limb darkening, given the strong irradiation in evidence.

The full model formulation is:
$\texttt{zxipcf}\times \texttt{tbFeO} \times [ \texttt{simplcut}(\texttt{kerrbb}+\texttt{mbknpo} \times \texttt{relxillCp} ) + \texttt{mbknpo} \times \texttt{xillverCp}] $.
Here, \texttt{mbknpo} is used to curtail unphysical runaway in the reflection spectrum at energies near and below the thermal disk's peak. This runaway occurs because the reflection model has been computed for a seed disk temperature of $kT_*=50$\;eV.   With \texttt{mbknpo}, we apply a break in the power-law shape of the reflection below a reference energy (typically several times the thermal-disk peak). Below this energy, the shape is forced to follow the low-energy tail of a multicolor disk (as one expects, given the thermal-disk photons that seed the Compton component). 

The photon index $\Gamma$ and electron temperatures $kT_{\rm e}$ of the reflection components are tied to corresponding settings in \texttt{simplcut}, which makes use of the \texttt{nthcomp} kernel.  The \texttt{xillverCp} component is assumed to originate with low ionization ($\log\xi=0$).  In order to allow for potential systematic uncertainties in the reflection model while also exploring potential misalignment between spin and orbital axes, we leave the inclination free in \texttt{relxillCp} (but not for \texttt{kerrbb}).  For analogous reasons, the spin is decoupled between components.

Owing to their short duration and high signal-to-noise, each \NICER\ good-time-interval (GTI) is fitted for mass accretion rate $\dot{m}$, coronal scattering fraction $f_{\rm sc}$, wind column $N_{\rm H}$, and reflection normalization.  A single set of those values is fitted across the long-exposure spectra from \IXPE, \NuSTAR, and \INTEGRAL. Each instrument is assigned a floating cross-normalization constant with respect to \NuSTAR's FPMB.  All other parameters are assumed to be invariant during an epoch.  For \IXPE\ and \AstroSAT\ instruments, the instrumental gain and energy zero-points are included as free parameters of the fit.  To account for instrumental residuals at Si-K in the NICER spectrum, we include a Gaussian absorption line at $1.74$~keV with $5$~eV width (Si K$\alpha$), and an edge  at $1.84$~keV.

\begin{figure}[t]
\includegraphics[clip=true,trim={0 0 0 0}, width=\linewidth]{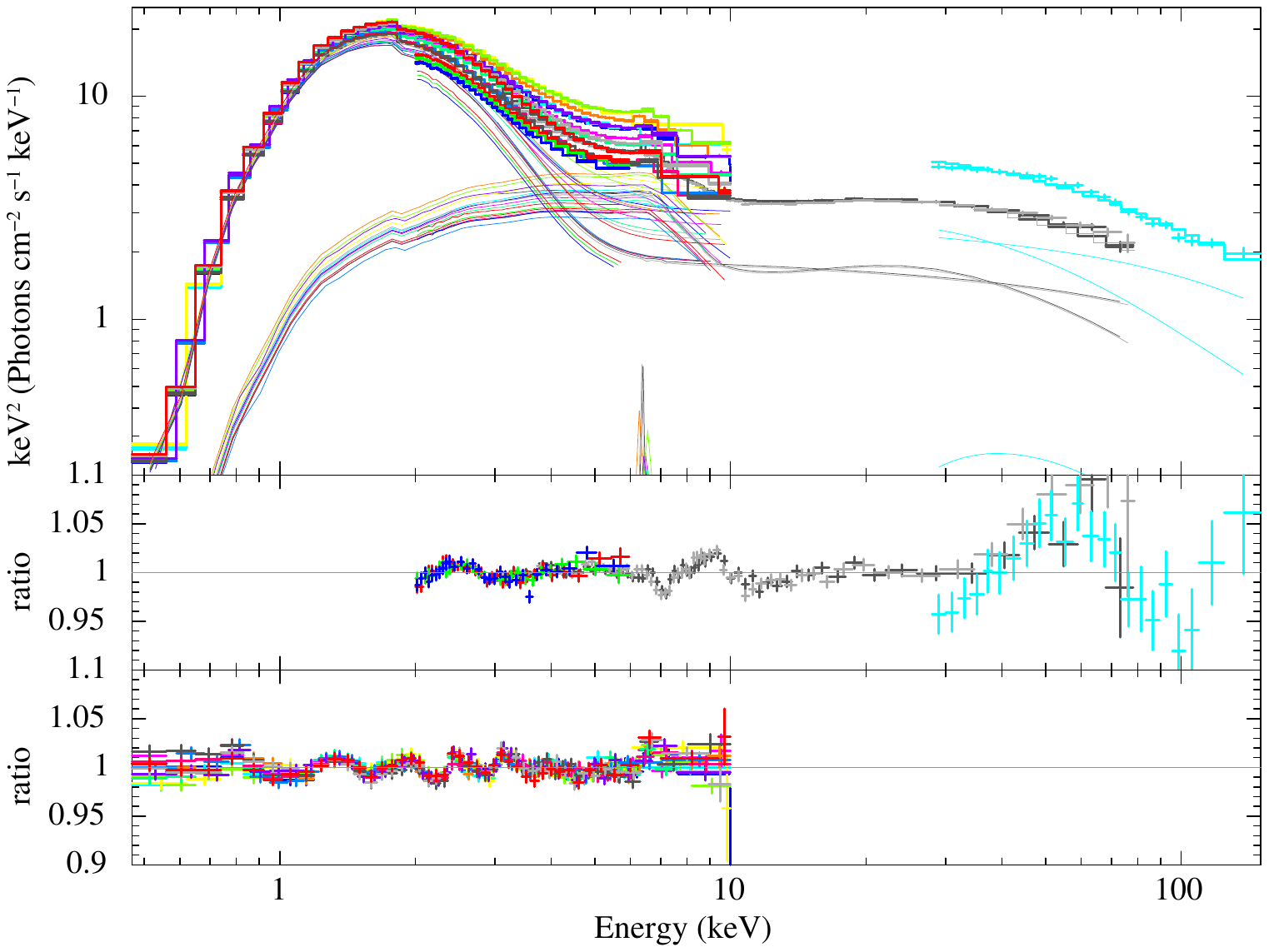}
\caption{The best-fitting comprehensive spectral model for Epoch 5, showing contributions from the Comptonized disk and reflection components. Each \NICER\ GTI contains 5--30 million counts; the \IXPE, \NuSTAR, and \INTEGRAL\ spectra contain $\sim$2--4 million counts apiece.
For clarity the ratios of data to best-fit model are split between the lower two panels. The middle panel shows \NuSTAR{} FMPA (dark grey), FMPB (light grey); \IXPE{} DU1 (red), DU2 (green), DU3 (blue); and \INTEGRAL{} IBIS (cyan).  The lower panel shows the ratio of the spectra from the different \nicer{} GTIs. The same color scheme is used for the unfolded data sets and model in the upper panel.}
\label{fig:specfit}
\end{figure}

Our comprehensive spectral fits are presented in Table~\ref{tab:results}, and illustrated for Epoch 5 in Figure~\ref{fig:specfit}. Because each spectrum contains millions of counts, many spectral bins are limited by systematic uncertainty which can arise from deficiencies either in the instrument calibration, or else in the spectral model. Although the fit statistics obtained are formally unacceptable, an additional systematic uncertainty from 0.4\%--0.8\% would result in a reduced $\chi^2$ of unity for each epoch, which we find eminently reasonable given typical calibration uncertainties (see, e.g., \citealt{Madsen_2017}).
We note that the break energy of the 
\texttt{mbkpno} reflection modifier is higher than might be expected for a disk with $kT_*\approx0.5$~keV, which we tentatively attribute to large gravitational redshift for this high-spin BH.  We additionally note that the inclination from the reflection modeling differs substantially among the epochs, with $i\approx30\degr$ for Epochs 4 and 5 in agreement with the binary orbital inclination from \citet{Orosz_2011}, whereas the higher inclination $i\approx40\degr$ from Epoch~3 more closely matches that obtained in \citet{Henric_ixpe_cygx1}.  The BH spins we find are all high. The continuum-fitting spin value is maximal for each fit, whereas the reflection spin is consistently high ($\gtrsim0.9$) but exhibits variance larger than the statistical uncertainty. We attribute these differences to systematic uncertainties in the model.  We note that if disk and reflection inclinations are linked, the inclination is driven to the disk value in Epochs 3 and 5. This linking doesn't have significant impact on the other model parameters, but produces a significantly worse fit.

For Epochs 3 and 5 (those with soft X-ray coverage), the fit was found to improve when including an additional thermal component, which fits with a temperature of $\sim1\textrm{--}1.5$~keV and a flux  $\sim10\%$ of the primary thermal component.  We speculate that such a component may originate from thermalized reflection returning radiation at the disk surface.  However, because other modeling systematics are of comparable magnitude,  we opt against including this additional component in our adopted model. 



\input{new_error_table}

\subsection{Empirical Polarimetric Fitting}

We fix the best-fitting spectral (and response-model) parameters and for each epoch attempt to fit the \IXPE\ Stokes-$Q$ and Stokes-$U$ data testing each of the three spectral components separately:  (i) the transmitted disk spectrum, (ii) the transmitted reflection spectrum, and (iii) the total Compton up-scattered spectrum.  Each component is assigned a constant PD and PA (\texttt{polconst}), and we assess which component(s) are capable of producing the observed signal.   In each case, the disk emission is insufficient, owing to its negligible contribution to high energies in particular.  Instead, either of the reflection or coronal components are capable of accounting for the observed signal.  Because the reflection signal is dominant over the coronal component, it need not be as highly polarized.  Using just the Stokes data from \IXPE, it is not possible to distinguish polarimetric contributions between the coronal and reflection components, which are similarly shaped over the 2--8 keV range. The observed polarization cannot empirically distinguished between the two components, and so a more detailed physical investigation is required, which we present in Section~\ref{sec:discuss}. Here, for this empirical analysis we consider alternate cases in which the polarization is solely attributed to one or the other component.

At 90\% confidence intervals for our preferred fits, for polarization to arise solely from the coronal component in Epochs 3, 4, and 5 would require PD=$17.8\%\pm3.3\%$, $8.0\%\pm3.6\%$, and $7.8\%\pm2.4\%$, respectively. The corresponding PA are $-16\degr\pm5\degr$, $-25\degr\pm13\degr$, and $-26\degr\pm9\degr$. 
If the polarization is instead attributed solely to reflection emission, the required PD=$10.9\%\pm2.0\%$, $4.3\%\pm1.9\%$, and $5.2\%\pm1.6\%$ respectively; the corresponding PA are $-16\degr\pm5\degr$, $-26\degr\pm13\degr$, and $-26\degr\pm9\degr$. 
We find that Epochs 4 and 5 are each marginally improved by allowing polarization of the disk emission, although the improvement is insignificant for Epoch 3.
For Epochs 4 and 5, the disk offers a marginal PD of $1.3\%\pm1.0\%$ and $1.3\%\pm1.0\%$, with PA of $-2\degr\pm25\degr$ and $-36\degr\pm26\degr$, respectively. 

Absent broadband X-ray spectral data for Epochs 1 and 2, we perform a simple analysis of these using a disk plus power-law model for the \IXPE\ data alone. At 90\% confidence, in Epochs 1 and 2, the power-law component has PD=$4.6\%\pm1.2\%$ and $4.5\%\pm0.7\%$, and PA=$-22\degr\pm8\degr$, $-21\degr\pm4\degr$, respectively.  The disk component has insignificant polarization in Epoch 2 but in Epoch 1 is marginally detected with PD=$3.2\%\pm3.0\%$ and PA=$53\degr\pm37\degr$.

\section{Discussion}
\label{sec:discuss}

\Cyg's soft state polarization properties are similar to those seen in other recent \IXPE\ observations of soft and SPL states, including \mbox{4U~1630$-$47} \citep{Ratheesh_2024, Rodriguez_2023}, \mbox{LMC X-1} \citep{Podgorny_2023}, \mbox{4U~1957+11} \citep{Marra_2024},  \mbox{LMC X-3} \citep{Svoboda_2024}, and \mbox{Swift J1727.8$-$1613} \citep{Svoboda_2024_J1727}.  \mbox{4U~1957+11} in particular shows similarly strong returning radiation as here, and in each system with sufficient signal, the PD increases with energy while the PA remains approximately fixed, in contrast to the classical expectation of large swings in PA and PD above the thermal peak for a BH with an electron-scattering disk atmosphere \citep{Connors_1980, Dovciak_2008, Schnittman_2009, Schnittman_2010}.

We investigated \cyg's spectro-polarization properties  using the general relativistic ray-tracing code \texttt{kerrC} \citep{KrawczynskiBeheshtipour+22,Henric_ixpe_cygx1}, adopting a wedge-shaped corona corotating with the accretion disk (see \cite{Poutanen_2023} and \cite{Dexter_2024} for an alternative explanation involving relativistic outflows).
{The \texttt{kerrC} code assumes a razor-thin accretion disk extending from the innermost stable circular orbit to 100 gravitational radii $r_{\rm g}\,=\,G\,M/c^2$. 
In \texttt{kerrC} the disk emits radiation polarized according to Chandrasekhar's classical results for a semi-infinite scattering atmosphere accounting for the reflection to all scattering orders \citep[][Equation 164 and Table XXV]{1960ratr.book.....C}, see \citet{2012ApJ...754..133K} and \citet{KrawczynskiBeheshtipour+22} for more details.  
\texttt{kerrC} can modulate the intensity of the reflected emission according to reflection radiative transport codes \citep[][and references therein]{2014ApJ...782...76G}, however we switch off that expensive capability here, owing to the high ionization of the disk.

As in the hard state analysis from \citet{Henric_ixpe_cygx1}, we use a wedge-shaped corona
extending from $r_{\rm ISCO}$ to a fixed 100\,$r_{\rm g}$ 
with a fixed half-opening angle of 10$^{\circ}$.  
We use a modified version of \texttt{kerrC} that implements a corona orbiting the symmetry axis at the position $(r,\theta,\phi)$ (Boyer Lindquist coordinates) with the Keplerian angular velocity evaluated in the equatorial plane at the radial coordinate $r_{\rm D}\,=\,\sin{\theta} r$.
In the original \texttt{kerrC} model, the corona orbits the symmetry axis with the  zero angular momentum observer (ZAMO) angular frequency. We find that the Kepler and ZAMO coronae give very similar flux and polarization energy spectra.
\texttt{kerrC} assumes a single temperature and vertical optical depth $\tau_{\rm C}$ for the entire corona (see the sketch in Fig.\,\ref{fig:kerrCplots}). \texttt{kerrC} assumes a 3D corona geometry, and so the effective optical depth varies spatially.  Note that perfect reflection off the disk increases the coronal flux at high energies substantially as photons back-scattered by the corona into the direction of the disk gain more energy than photons forward-scattered into the direction of the observer. 
The disk can reflect these higher-energy photons towards the observer. The reflecting disk furthermore increases the effective optical depth \citep[see also][]{1993ApJ...413..680H}. 
The polarization change in coronal scatterings is effected in the electron rest frame using Fano's relativistic scattering matrix \citep{1957RvMP...29...74F,2017ApJ...850...14B}.

We fixed the BH mass, distance, spin, and accretion rate to the \texttt{kerrbb} values assumed or fitted in Table~\ref{tab:results}. We tested the binary-orbital inclination, the reflection inclinations from Table~\ref{tab:results}, and several other reference values. 
The {\em vertical} coronal optical depth $\tau_{\rm C}$ ($\sim$ 0.007 for Epoch 5) and the corona electron temperature $kT_{\rm e}$ ($\sim$ 250 keV) were obtained from an eyeball fit of the \NICER\ and \NUSTAR\ spectral data. 
The vertical coronal optical depth for Epoch 5 
was significantly larger than for Epochs 3 or 4 ($\tau_{\rm C} \sim 0.002$), 
in-line with the trend of $f_{\rm sc}$ in the spectral fits. 
A comparison between the \texttt{kerrC} polarization prediction and the \IXPE\ data is displayed in Figure~\ref{fig:kerrCplots}. 
We find that for \texttt{kerrC} to match the high PD values observed, high spins of $\spin\gtrsim0.96$ are required. At these high spins, the inner edge of the disk is very close to the event horizon, and so due to gravitational lensing from the strong spacetime curvature a large fraction of photons, both thermal and reflection, return to the disk (e.g., \citealt{Dauser_2022}).  The latter is most important at the higher energies in \IXPE's bandpass. These reflect off the disk and generate high PD \citep{Schnittman_2009,Taverna2020,KrawczynskiBeheshtipour+22}.
The PDs are slightly underpredicted for an inclination of $i\,=\,27^{\circ}$ and slightly overpredicted for $i\,=\,40^{\circ}$.
For the $i\,=\,27^{\circ}$ model the blue dotted line shows the result when reflection photons are omitted, the resulting deficit in polarization highlights the large contribution from these reflected photons. 
The orange dashed line shows the impact of removing the corona from the $i\,=\,40^{\circ}$ model.  Here, the disk with its reflected returning radiation produces even higher PD than from the disk-corona model.  The high polarization for this case was verified using \texttt{kynbbrr} 
 \citep{Taverna2020}.
Additional exploration with \texttt{kerrC} reveals that the net PD and PA values are insensitive to the coronal temperature $kT_{\rm e}$, as reflection dominates the polarization signature. 
We note that Fe-K$\alpha$ emission is expected to reduce the PD in the 6--8\,keV energy band.
The effect on the PD is, however, one order of magnitude smaller than the {\it IXPE} measurement error. 

We separately investigated predictions for the disk plus slab-corona model using the polarimetric Comptonization code \texttt{compps} \citep{compps,Veledina2022}.
Ray-tracing was performed with the  code \textsc{artpol} based on analytical results \citep{Loktev2022,Loktev2024}.   
A slab geometry was adopted with 
$\tau=0.2$ to match the slope of the power law observed at high energies. 
A purely Maxwellian electron distribution with $kT_{\rm e}=92$~keV was adopted and the spin set to the maximum value allowed by \textsc{artpol} ($\spin=0.94$).  The associated Compton-component polarization was very low. Specifically, the predicted PDs are substantially lower than the observed values for inclinations between  $30\degr$ and $45\degr$ (e.g., $<1\%$ in the \IXPE\ bandpass at $i=30\degr$).

In \texttt{kerrC} and \texttt{kynbbrr}, the transition from dominance of direct emission to reflection emission produces a $\sim90^{\circ}$ PA swing around 0.5\,keV, whereas absent reflection, from Comptonization in the corona, such a swing would occur near 5\,keV energy in the transition from disk-dominated to the corona-dominated energy band. 
As can be seen in Fig.~\ref{fig:kerrCplots}, when reflection is omitted from \texttt{kerrC}, the predicted the Compton-component polarization are very low, 
in good agreement with the \texttt{compps} results.
Higher PDs could be generated by combining lower electron temperatures with higher optical depths \citep[e.g.,][]{ST85}, or with a hybrid (thermal and non-thermal) electron distribution  \citep{Gierlinski_1999}, with nonthermal electrons expected to dominate the Comptonization tail at highest energies.

In \Cyg's  \ixpe+\NICER+\NuSTAR\ hard-state analysis by \citet{Henric_ixpe_cygx1}, the PA was measured to be stable with energy and in alignment with both the radio jets orientation and the intrinsic PA in optical \citep{Kravtsov_2023}.  
At the same time, the PD was found to increase approximately linearly with energy in the best fit, much as here.  

Figure~\ref{fig:kTscl} shows an intriguing trend of the PD increasing with the energy in units of its natural scale, the disk's temperature. In the soft state, our best model attributes the PD increase with energy to the growing dominance of returning-radiation induced reflection. 
The hard state fit was explained as PD rising owing to the increasing number of scatterings experienced by higher-energy photons.
The appearance of a common trend produced by the hard and soft state PD energy spectra is intriguing although we note it may simply be coincidental.

\begin{figure}[t]
\includegraphics[width=0.332\linewidth]{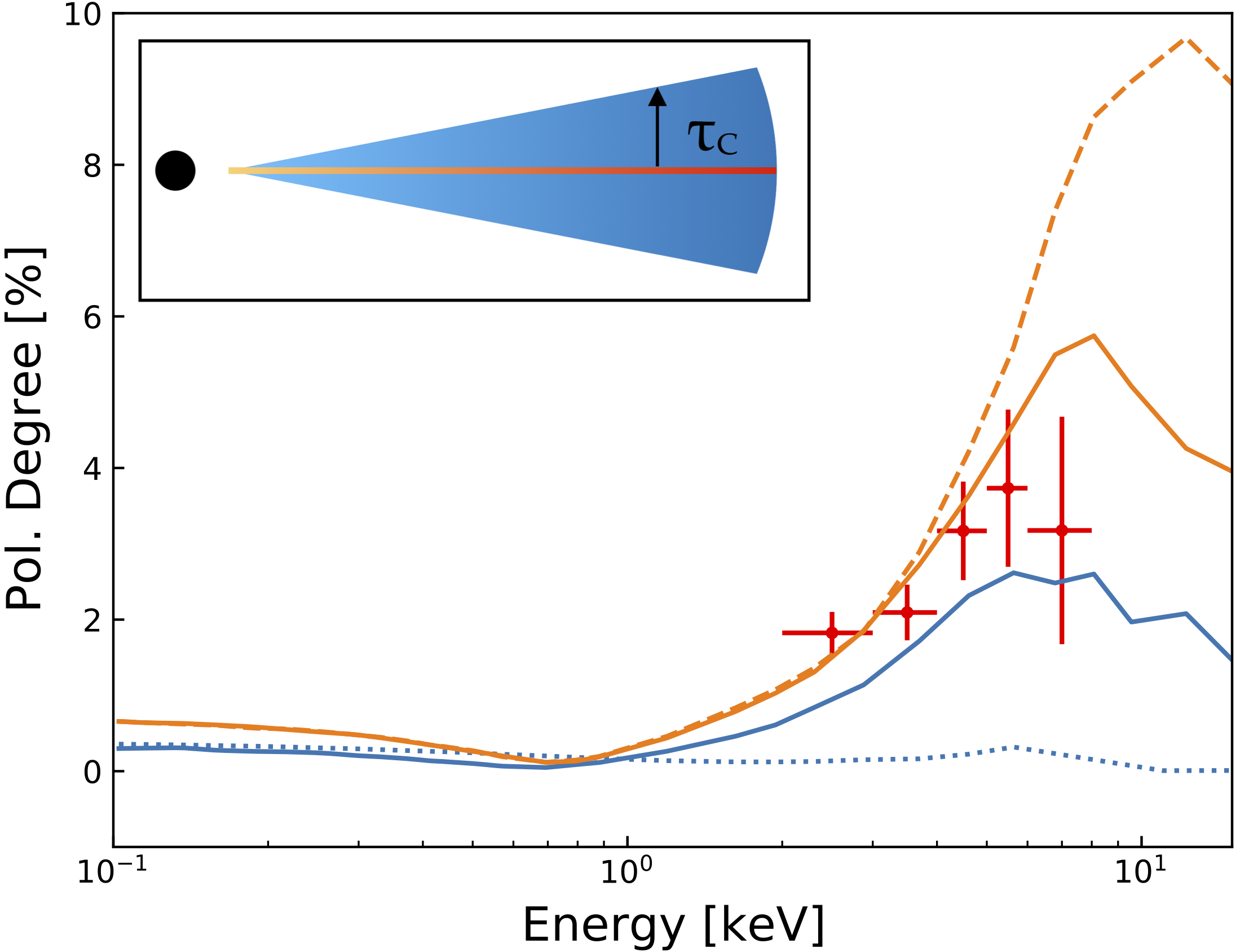}
\includegraphics[width=0.332\linewidth]{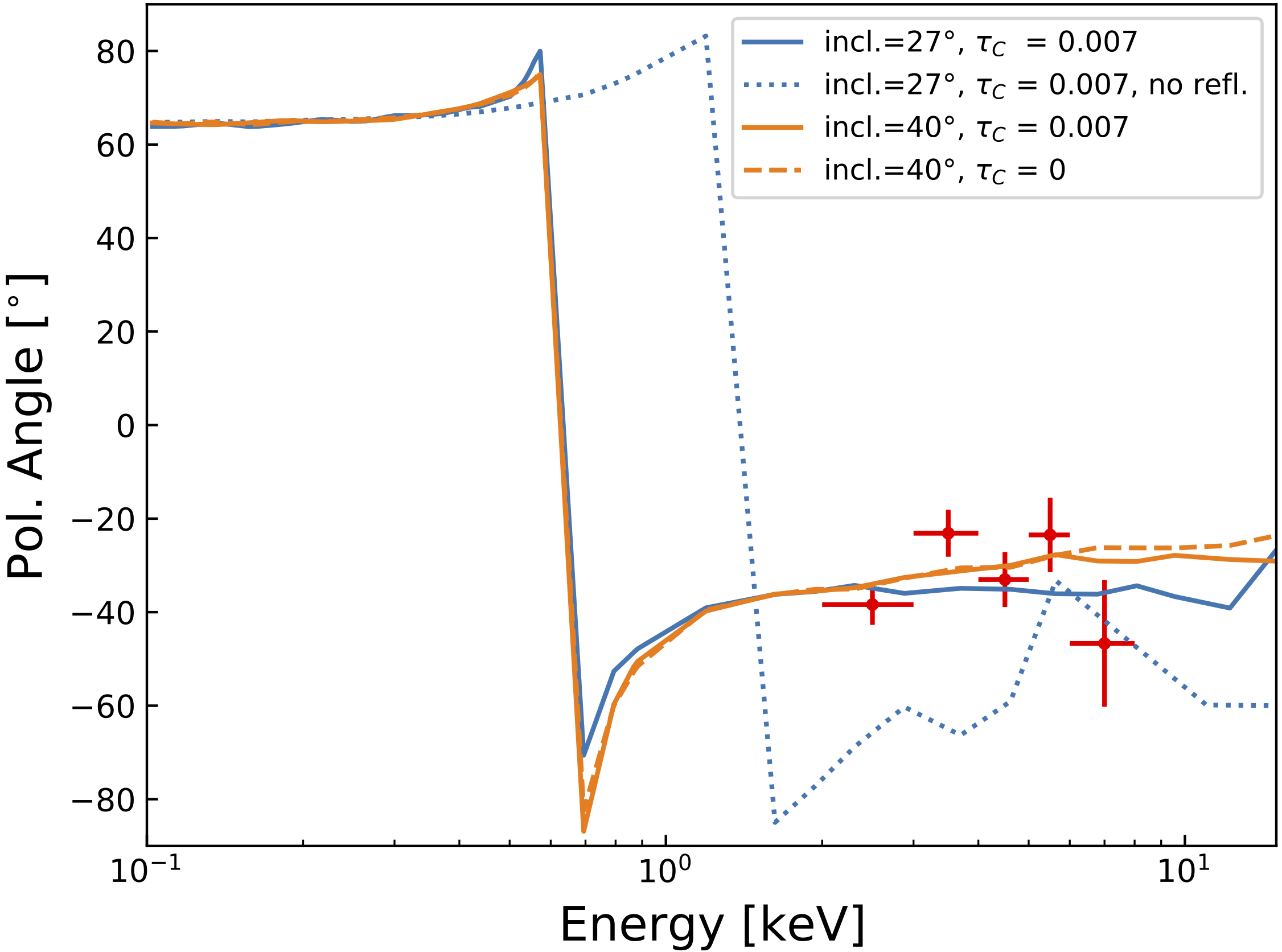}
\includegraphics[width=0.332\linewidth]{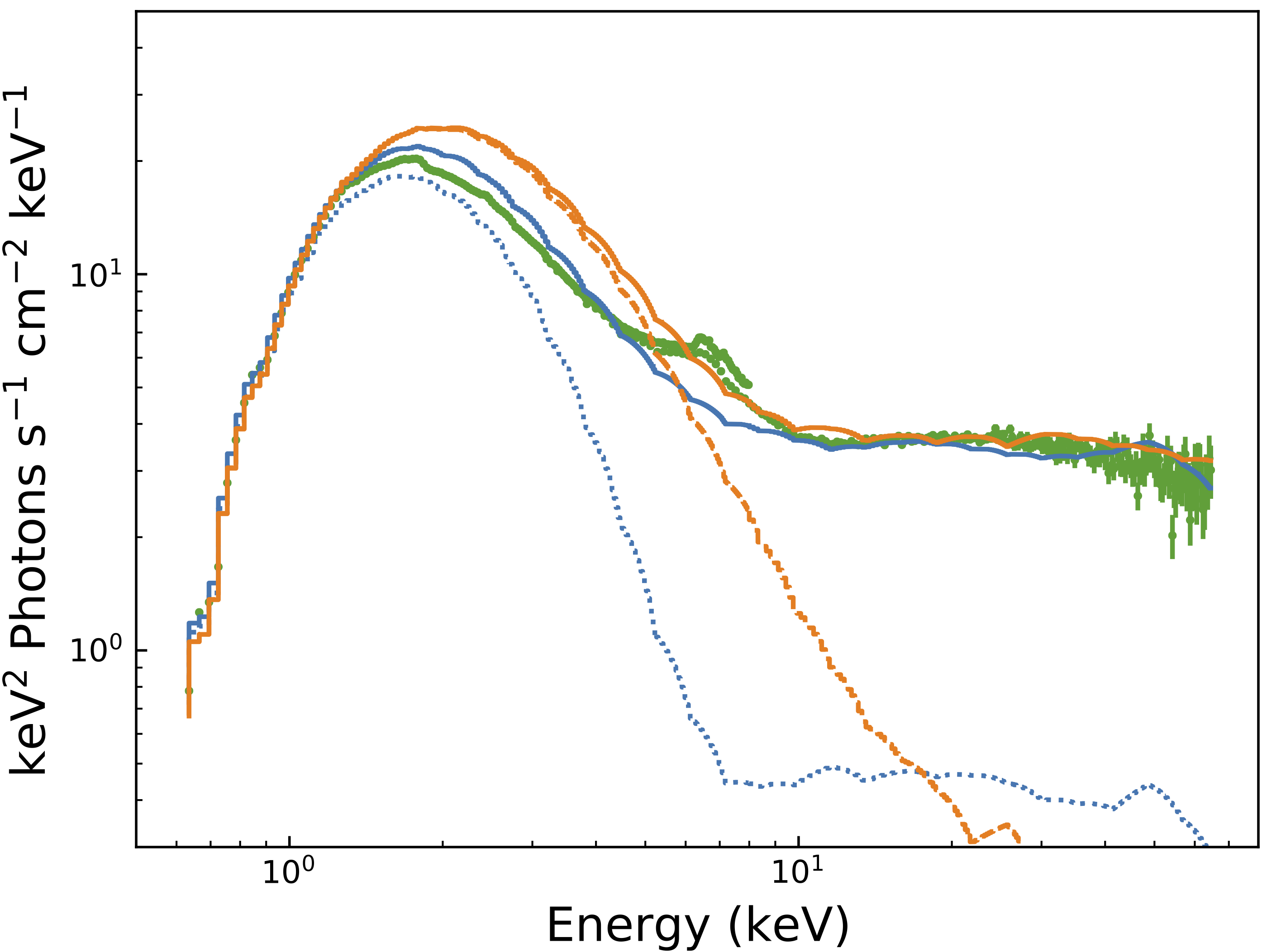}
\caption{\texttt{kerrC} simulations of the PD (left), PA (middle), and spectrum (right) expected from a standard geometrically thin optically thick accretion disk of a rapidly spinning black hole ($\spin=0.998$) with a hot wedge-shaped corona (illustrated in the left panel inset). 
We show results for inclinations $i\,=\,27^{\circ}$ (close to the orbital inclination, blue lines) and a higher inclination $i\,=\,40\degr$ (orange lines). 
We furthermore demonstrate the effect of removing the corona but not the returning emission (dashed orange line), and of removing all reflected emission (blue dotted line).
The polarization signal is clearly dominated by the reflected emission. 
The {\it IXPE} polarization data for Epoch 5 (red data points) support high spin, owing to the impact of the reflected emission.Note the polarization swing at $\sim$0.5\,keV for all models with disk reflection.
Very similar results are obtained for Epochs~3 and~4.   Green points in the right panel depict the \NICER\ and \NuSTAR\ data.}
\label{fig:kerrCplots}
\end{figure}

\begin{figure}[t]
\begin{center}
\includegraphics[clip=true,trim={0 0 0 0}, 
width=0.6\linewidth]{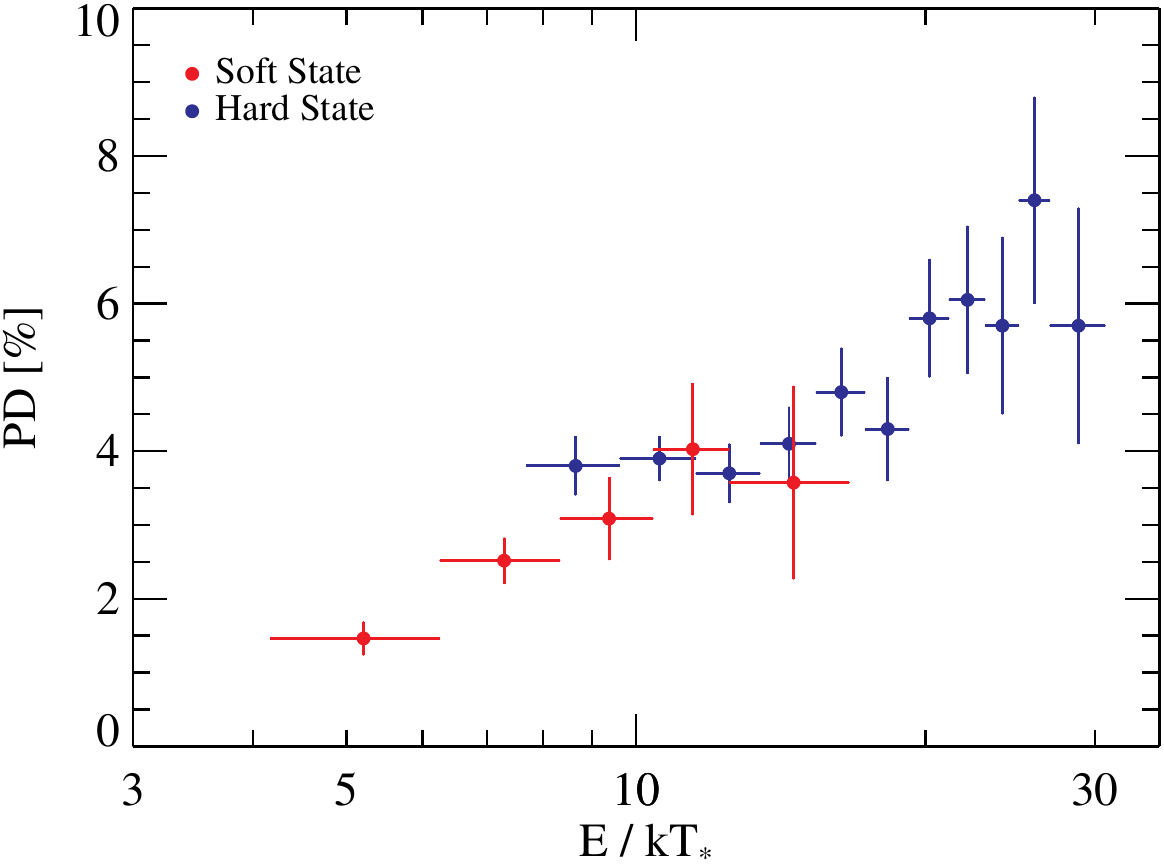}
\caption{\IXPE\ PD for the composite of hard vs soft states, where the energy-bins have been scaled with respect to the disk temperature $kT_*$.  
The consistent behavior in evidence is suggestive of commonality between hard and soft states, despite radically-different spectral-timing characteristics.
}
\label{fig:kTscl}
\end{center}
\end{figure}

\section{Conclusions}
\label{sec:conc}

We present the first \IXPE\ polarimetric observations of \cyg's soft state.  
The soft state exhibits weaker polarization (PD$\approx2\%$) compared to the hard state, but 
in every other respect the polarimetric properties of both states are strikingly similar, including a constant PA$\approx-26\degr$ (a direction parallel to \Cyg's jet) and a rising PD with energy.  
 This commonality is bolstered by their adherence to a single track of PD versus energy when normalizing the \IXPE\ energy bins by the observed disk temperature (Figure~\ref{fig:kTscl}). At the same time, our best model attributes the polarization signatures in hard and soft states to different dominant effects.  Whereas in the \IXPE\ band, hard state polarization was explained as originating from multiply-scattered photons in the corona, the soft state polarization is explained as the result of returning-radiation associated with the reflection emission, a consequence of the high spin and strong-gravitational lensing for the inner disk. 
 
In addition to the polarimetric analysis, accompanying multiwavelength data in radio, optical, and X-ray are also presented.  Broadband X-ray spectral data span $\sim 0.5-500$~keV.  The associated spectral fits use tens of millions of counts, anchoring the thermal disk, coronal, and reflection contributions.
Spectropolarimetric modeling of these data with the wedge-shaped coronal geometry used to fit the hard state (with \texttt{kerrC}) also proved successful at describing the soft state as well.  The polarimetric results constrain a high spin ($\spin\gtrsim0.96$) and allow up to a $\lesssim10\degr$ more-inclined spin axis compared to the binary orbit.

\begin{acknowledgments}

 The Imaging X-ray Polarimetry Explorer (\IXPE) is a joint US and Italian mission.  The US contribution is supported by the National Aeronautics and Space Administration (NASA) and led and managed by its Marshall Space Flight Center (MSFC), with industry partner Ball Aerospace (contract NNM15AA18C).  The Italian contribution is supported by the Italian Space Agency (Agenzia Spaziale Italiana, ASI) through contract ASI-OHBI-2022-13-I.0, agreements ASI-INAF-2022-19-HH.0 and ASI-INFN-2017.13-H0, and its Space Science Data Center (SSDC) with agreements ASI-INAF-2022-14-HH.0 and ASI-INFN 2021-43-HH.0, and by the Istituto Nazionale di Astrofisica (INAF) and the Istituto Nazionale di Fisica Nucleare (INFN) in Italy.  This research used data products provided by the \IXPE\ Team (MSFC, SSDC, INAF, and INFN) and distributed with additional software tools by the High-Energy Astrophysics Science Archive Research Center (HEASARC), at NASA Goddard Space Flight Center (GSFC). 

Part of the radio data was obtained at RATAN-600 SAO RAS, with the  data processing supported by the   Ministry of Science and Higher Education of the Russian Federation grant No.~075-15-2022-262 (13.MNPMU.21.0003).
This work makes use of data from the AstroSat mission of the Indian Space Research Organisation (ISRO), archived at Indian Space Science Data Centre (ISSDC). 
The article has used data from the SXT and the LAXPC developed at TIFR, Mumbai, and the AstroSat POCs at TIFR are thanked for verifying and releasing the data via the ISSDC data archive and providing the necessary software tools.
The results presented include work supported by Tamkeen under the NYU Abu Dhabi Research Institute grant CASS.

JFS acknowledges support from NASA contract NAS8-03060, and NuSTAR General Observer Program 80NSSC23K1660.
EN acknowledges support from NASA theory grant 80NSSC20K0540.
MD, JS, JPod and VKar thank GACR project 21-06825X for the support and institutional support from RVO:67985815.
AV thanks the Academy of Finland grant 355672 for support. 
AAZ acknowledges support from the Polish National Science Center
grants 2019/35/B/ST9/03944 and 2023/48/Q/ST9/00138, and from the
Copernicus Academy grant CBMK/01/24.
P-OP and MP acknowledge support from the High Energy National Programme (PNHE) of Centre national de la recherche scientifique (CNRS) and from the French space agency (CNES).  IL was supported by
the NASA Postdoctoral Program at the Marshall Space Flight Center,
administered by Oak Ridge Associated Universities under contract with
NASA. FMu, ECo, ADM, RF, PSo, SF, FLM  are partially supported by MAECI with grant CN24GR08 ``GRBAXP: Guangxi-Rome Bilateral Agreement for X-ray Polarimetry in Astrophysics''.  We thank the anonymous referee for their stimulating feedback on our investigation.

\end{acknowledgments}

\facilities{IXPE, NuSTAR, NICER, AstroSat, INTEGRAL, RATAN, AMI, LCO, Perek Telescope}

\software{ixpeobssim, heasoft}

\appendix
\label{appendix}

\section{Data Description}
\label{sec:app:data}
\subsection{IXPE}
\label{app:ixpe}

\IXPE\ Epochs 1 through 5 correspond to \IXPE\ Observation IDs 02008201, 02008301, 02008401, 02008501, and 02008601, respectively.
During these observations of \cyg, owing to the source brightness, one detector (DU3) exhibited an excess of events at high energies, resulting in a hard tail. Such events are induced by instances of residual charge in the detector's region of interest (ROI) producing unphysical tracks. These events have a signature similar to those of background events, and so can be removed by background screening, see \cite{DiMarco_2023} for a complete discussion.   These were correspondingly removed via  background screening using the \texttt{filter\_background.py} software.\footnote{\href{https://github.com/aledimarco/IXPE-background}{https://github.com/aledimarco/IXPE-background}}

From cleaned level-2 event data for each gas pixel detector (GPD), $80\arcsec$ and $60\arcsec$ apertures were used to extract source data for polarimetric and spectroscopic analysis, respectively, using the  \texttt{ixpeobssim} software suite \citep{ixpeobssim}. A background region was extracted from an annulus with inner and outer radii of $150\arcsec$ and $310\arcsec$, centered on the source position.  
Data were extracted into ``polarization cube'' structures, allowing ready data slicing by detector, time, energy, etc.   

We cross-checked these extractions by performing a standard analysis using {\sc xselect}, using the effective event-number weighting (STOKES=``NEFF'') to produce equivalent products in Stokes $I$, $Q$, and $U$ parameters.  In both instances, standard weighting with track ellipticity scaled to the 0.75 power was adopted, and the corresponding response files (v2 of the rmf and v5 of the arf) were used.  Exposure times were corrected for $\lesssim20\%$ detector deadtime.  The \IXPE\ Stokes $Q$ and Stokes $U$ data were analyzed in the 2--8 keV range, while the Stokes $I$ data were analysed in the 2--6 keV range due to flux calibration uncertainty.

\subsection{NuSTAR}
\NuSTAR\ \citep{NuSTAR} observed \cyg\ during each of Epochs 3, 4, and 5 (Observation IDs 80902318002, 80902318004, and 80902318006, respectively).   A portion of the data collected during the latter two epochs was never relayed owing to problems with ground station contacts ($\sim5.5$~ks for Epoch 4 and $\sim3.5$~ks for Epoch 5).   \Cyg\ was sufficiently bright that those data were overwritten onboard and were not recoverable.  The three observations yielded exposure time of 13.8, 9.5, and 10.5~ks, respectively.  Each observation was reduced following standard procedures for bright sources, including the modified {\sc nufilter} condition ``STATUS==b0000xxx00xxxx000" for producing clean level-2 data. Source events were extracted from a $100\arcsec$ radius region centered on the source peak in each focal plane module (FPM), and in each case the background was obtained from a box 750$\arcsec\times120\arcsec$ at the detector edge.  The spectral data were optimally binned via {\sc ftgrouppha} \citep{Kaastra_Bleeker_2016}, deadtime corrected, and analyzed from 3--79~keV.  As the result of a tear in the FPMA thermal blanket \citep{Madsen2020}, following the guidance of the \NuSTAR\ helpdesk, spectral analysis made use of the empirical correction table \texttt{NuMLIv1.mod} which was used to adjust only FPMA, in order to  account for its impact on the low-energy response.

\subsection{NICER}
\NICER\  \citep{NICER} observed \Cyg\ during Epochs 3 (Observation IDs 6643010101 and 6643010102; 8 useful GTIs in total) and 5 (Observation IDs 6643010103 and 6643010104; 13 useful GTIs in total).  GTIs were of typical duration $\sim 1$~ks and were generally  separated by one orbit of the International Space Station ($\sim 90$ minutes), but sometimes separated by multiple orbits.  All of these observations took place after the discovery of an optical light leak\footnote{\url{https://www.nasa.gov/feature/goddard/2023/nicer-status-update}} caused by a damaged thermal shield on one of the detectors.  During ISS daytime, the leak contributes additional noise at low energies and also can produce telemetry saturation from detector reset events.  The daytime data were found to suffer significant packet losses and so data were screened to select only dark conditions (filter setting ``sunshine==0'').  Aside from this requirement, data were subject to standard level-2 processing and filtering using {\sc nicerdas}-v10, and extracted per continuous GTI.  Any GTI less than 60\,s was discarded.  A total of 6.8~ks and 14.5~ks were produced for the two epochs, in 8 and 13 GTIs, respectively (Table~\ref{tab:observations}).  For each GTI, individual detector behavior was screened against the instrument ensemble for each of undershoot (reset) event rates, overshoot (particle) event rates, and X-ray event rates, with any 10-$\sigma$ equivalent outlier detector flagged and excised from the ensemble.  FPM 63 was flagged and removed in this way owing to an excess undershoot rate, which was seen in all GTIs.  FPM 55 had been turned off during observations.  The remaining 50 detectors were combined for all subsequent analysis.  The \NICER\ count rate varied between 20,000\;s$^{-1}$ and 24,000\;s$^{-1}$ (52-FPM equivalent), approximately twice the rate of the Crab.  We computed backgrounds for each \NICER\ GTI using the mission-recommended \texttt{scorpeon}\footnote{\url{https://heasarc.gsfc.nasa.gov/docs/nicer/analysis_threads/scorpeon-xspec/}} and {\sc 3c50} \citep{Remillard_3C50}  background models.  These were negligible in comparison with the data in each case, and the background models were in close agreement. The exposure times were adjusted for $\sim1\%$ detector deadtime.  \NICER\ spectral data were binned to oversample the detector energy resolution by a factor $\sim 2$ and analyzed over the range 0.5--11 keV with a 1\% systematic error included.

\subsection{INTEGRAL}
\INTEGRAL\ observed \cyg\ during Epochs 4 and 5, from 2023 June 13 22:15:42.865 UTC to 2023 June 15 14:51:31.971 UTC (\INTEGRAL\ revolution 2651) and from 2023 June 20 00:00:09.268 UTC to 2023 June 20 15:04:10.319 UTC (\INTEGRAL\ revolution 2653). 
We consider all \INTEGRAL\ individual pointings or science windows (scws) during these two periods.  To probe the source behavior in the hard X-rays, we make use of data from the first detector layer of the Imager on Board the INTEGRAL Satellite (IBIS), the INTEGRAL Soft Gamma-ray Imager (ISGRI), which is sensitive between $\sim20$ and $\sim600$ keV \citep{Lebrun2003}. Data were reduced with version 11.1 of the Off-line Scientific Analysis (OSA) software following standard procedures.\footnote{\url{https://www.isdc.unige.ch/integral/download/osa/doc/11.0/osa_um_ibis/Cookbook.html}}
For each scw, a sky model was constructed, and the sky image and source count rates were reconstructed by deconvolving the shadowgrams projected onto the detector plane.
Spectra were extracted using 40 logarithmically spaced channels between 20 keV and 1000 keV. Response matrixes were automatically generated running the OSA 11.1 spectral extraction. Subsequently, the \texttt{spe\_pick} tool was employed to create stacked spectra for each distinct epoch, with the addition of a 2\% systematic uncertainty, in accordance with the specifications outlined in the IBIS user manual.
During the spectral fitting, a hard feature $\gtrsim200$~keV dominated the flux.  As this component was sufficiently far from the \IXPE{} band of interest, we assume this feature does not significantly effect the spectrum at lower energies, thus the spectral fits presented in Section~\ref{sec:results} only considered the \INTEGRAL\ data below 150~keV.

\subsection{AstroSat}
\label{x:astrosat}

AstroSat \citep{Singh_2014} observed \cyg\ during Epoch 3 from 2023 May 24, 19:11:50 UT to 2023 May 25, 11:55:00 UT (Observation ID: T05\_105T01\_9000005662). The primary instrument for the observation was the Soft X-ray Telescope (SXT, \citealt{Singh_2016,Singh_2017}) operating in Fast window (FW) mode. The Large Area X-ray Proportional Counter (LAXPC, \citealt{Yadav_2016, Yadav_2017}) also observed the source simultaneously in Event Analysis (EA) mode.

We procured the level-2 data for SXT (as processed by the Payload Operation Centre, POC) and extracted standard products (i.e. light curve and the spectra) for individual AstroSat orbits using {\sc xselect}. We used an annular region with an inner radius of 3\arcmin\ and outer radius of 5\arcmin\ to mitigate pile-up effects in the extracted products. The spectra were extracted for individual AstroSat orbits and merged via {\sc addspec}. We used standard response and background files provided by the SXT POC, and modified the ancillary response file to correct for the annular region adopted.  For spectral modeling, we fit the energy range 0.8--7.0 keV \citep{Bhargava_2023} and adopt a 3\% systematic error.

LAXPC level 1 data were processed using \textsc{LAXPCsoftware22Aug15}  \citep{Antia_2021, Misra_2021}. We obtain the spectrum, light curve, background spectrum, and responses using pipeline tools. LAXPC data are fitted from 3--35 keV, beyond which the spectrum is background dominated. We include 3\% systematic error in our analysis to mitigate residual uncertainties in the response \citep{Bhargava_2022}.  We use data from only one proportional counter unit (LAXPC20; \citealt{Antia_2021}) as LAXPC30 ceased operation early in the mission due to gas leakage, and LAXPC10 has presented abnormal gain variations.

\subsection{LCO}
\label{x:LCO}

Optical monitoring of \cyg\ was performed with the LCO 1m robotic telescopes located in McDonald Observatory (Texas, USA) and with the Teide Observatory (Tenerife, Spain), from 2023 June 02 (MJD 60097.34) to 2023 July 02 (MJD 60127.93), using B, V, r$^{\prime}$ and i$^{\prime}$ filters. Due to the brightness of the source, all the observations had 2\,s exposure times to avoid saturating the instrument.  The acquired images were processed and analyzed by the XB-NEWS pipeline (see \citealt{Russell_2019} and \citealt{Goodwin_2020}), carrying out the following tasks: 
\begin{enumerate}\vspace{-1.5ex}
\item Download fully-reduced images from the LCO database (i.e., bias, dark, and flat-field corrected images).\vspace{-1.5ex}
\item Reject any images of poor quality. \vspace{-1.5ex}
\item Perform astrometry using Gaia DR2 positions. \vspace{-1.5ex}
\item Carry out multi-aperture photometry (MAP; \citealt{Stetson_1990}).\vspace{-1.5ex}
\item Solve for photometric zero-point offsets between epochs \citep{Bramich_Fredling_2012}.\vspace{-1.5ex}
\item Flux-calibrate the photometry using the ATLAS-REFCAT2 catalog \citep{Tonry_2018}. \vspace{-0.5ex}
\end{enumerate}

If the target is not detected in an image above the defined detection threshold (a very unlikely prospect for \Cyg), XB-NEWS performs forced MAP at the target coordinates. If a forced MAP was performed we reject any with an uncertainty above 0.25 mag. After XB-NEWS data processing, a total of 17, 15, 14 and 15 data points in  B, V, r$^{\prime}$ and i$^{\prime}$, respectively, are obtained, spanning the latter half of the \IXPE\ campaign.

\subsection{RATAN}
\label{x:RATAN}
We have carried out observations of \Cyg\ with the RATAN-600 radio telescope at 4.7 GHz and 11.2 GHz from 16 to 24 June 2023 using the ``Southern Sector and Flat mirror'' antenna. The sensitivity of such measurements are about 3--10 mJy per beam. Thus \Cyg\ was undetected most of the time, with upper limits and detections presented in Figure~\ref{fig:mw}. Previous monitoring observations of \cyg\ have shown typical flux variations in the vicinity of 10--30 mJy at 4.7 GHz. Calibration was performed using quasar 3C\,48, adopting a brightness of 5.8 and 3.42 Jy at 4.7 and 8.2~GHz, respectively, according to the flux density scale by \citet{Ott_1994}.

\subsection{AMI}
\label{x:AMI}
\Cyg\ was observed 44 times in 2023 May and June
with Arcminute Microkelvin Imager (AMI) Large Array
\citep{2008MNRAS.391.1545Z,2018MNRAS.475.5677H} at 15.5~GHz.
The observations were typically $\sim 25$~min, with two 10-min
scans on \cyg\ interleaved between short observations
of a nearby compact source. The flux density scale of the
observations was set by using daily short observations
of 3C~286, and the interleaved calibrator observations were used to calibrate antenna based amplitude and phase variations during
the observations. The observations covered a 5~GHz bandwidth, 
of a single linear polarization, Stokes $I-Q$.

\subsection{Perek Telescope}
\label{x:perek}
\Cyg\ was monitored in optical with the 2m Perek Telescope located in the Ondřejov Observatory in the Czech Republic. The Ondřejov Echelle Spectrograph (OES) and the Single Order Spectrograph (CCD700) observed the source for an hour of exposure in the V-band.   Observations were first conducted just prior to the \IXPE\ campaign on 2023 April 27 at 00:30:34 UTC, with later observations near Epoch 1 on 2023 May 04 at 24:18:4 UTC, during Epoch 3 on 2023 May 25 at 00:45:34 UTC, shortly before Epoch 5 on 2023 June 18 at 24:22:8 UTC, and then a week after the \IXPE\ campaign on 2023 June 29 at 21:39:44 UTC.  In the H$\mathrm{\alpha}$ spectral region (6562 $\mathrm{\mathring{A}}$), the OES achieves a high spectral resolution of 40000, while the CCD700 only reaches 13000. The CCD700 is therefore principally used for calibration. For more technical information about the spectrographs, see \citet{OES} and \citet{Kabath_2020}. We reduced and processed the spectra using \texttt{IRAF} software \citep{Tody_1986SPIE..627..733T,Tody_1993ASPC...52..173T}. For the OES spectra, we use a semi-automatic reduction pipeline (see \citealt{cabezas_2023_10024183}). This pipeline includes wavelength and heliocentric calibration and continuum normalization. The disentangling method of \citet{Hadrava_2009} combines the optical spectra of the source at different orbital periods to measure the radial velocities and the orbital parameters. The H$\alpha$ P-Cygni profile is then isolated and the strength factor of the line is calculated with respect to the continuum.  

\bibliography{biblio}

\end{document}

%% file: new_error_table.tex
\startlongtable
\begin{deluxetable}{rlllll}
\tabletypesize{\footnotesize}
\tablecolumns{5}
\tablewidth{0pc}
\tablecaption{X-ray Spectral Fits}
\tablehead{Component &                           Variable &                 Unit &                           Epoch 3 &                         Epoch 4 &                         Epoch 5 }
\startdata 
TBfeo          & $\nh$                                      &   $10^{22}$cm$^{-2}$ &      $0.7501^{+0.0021}_{-0.0005}$ &                      $0.75$ (f) &      $0.7487^{+0.0009}_{-0.0008}$ \\
               & O                                        &                      &         $1.072^{+0.005}_{-0.003}$ &                      $1.08$ (f) &         $1.080^{+0.004}_{-0.002}$ \\
               & Fe                                       &                      &                     $0.46\pm0.02$ &                      $0.42$ (f) &                    $0.41\pm0.02$  \\
zxipcf         & $\nh$                                      &   $10^{22}$cm$^{-2}$ &                        $2.76$ (t) &                      $3.03$ (f) &                        $3.21$ (t) \\
               & $\log\xi$                                &                      &                     $2.30\pm0.02$ &                      $2.10$ (f) &                     $2.07\pm0.02$ \\
               & CvrFract                                 &                      &      $0.258^{+0.006}_{-0.011}$ &                      $0.22$ (f) &         $0.246^{+0.010}_{-0.007}$ \\
simplcut       & $\Gamma$                                 &                      &         $1.988^{+0.002}_{-0.003}$ &       $2.090^{+0.009}_{-0.005}$ &         $2.090^{+0.004}_{-0.002}$ \\
               & $f_{\rm sc}$                             &                      &      $0.0384^{+0.0027}_{-0.0007}$ &       $0.057^{+0.003}_{-0.002}$ &                   $0.076\pm0.002$ \\
               & ${kT}_\text{e}$\tablenotemark{a} &                  keV &         $500^{\text{(p)}}_{-140}$ &       $500^{\text{(p)}}_{-110}$ &         $500^{\text{(p)}}_{-120}$ \\
mbkpno         & B                                        &                  keV &           $3.87^{+0.05}_{-0.03}$  &          $3.80^{+0.07}_{-0.05}$ &                     $2.55\pm0.04$ \\
relxillCP      & $i$                                     &                  deg &                      $32.7\pm0.6$ &                    $30.7\pm1.2$ &                      $29.1\pm0.4$ \\
               & $\spin$                                       &                      &         $0.931^{+0.015}_{-0.004}$ &          $0.89^{+0.07}_{-0.04}$ &         $0.925^{+0.003}_{-0.005}$ \\
               & Rbr                                      &  ${r}_\text{G}$ &            $4.03^{+0.08}_{-0.17}$ &             $7.9^{+1.6}_{-1.2}$ &            $3.90^{+0.06}_{-0.08}$ \\
               & Index1                                   &                      &               $6.6^{+0.1}_{-0.1}$ &                     $5.0\pm0.3$ &               $7.5^{+0.1}_{-0.2}$ \\
               & Index2                                   &                      &            $3.13^{+0.02}_{-0.03}$ &             $2.5^{+0.1}_{-0.3}$ &         $2.905^{+0.013}_{-0.014}$ \\
               & $\log\xi$                                &                      &         $3.341^{+0.004}_{-0.015}$ &          $3.81^{+0.05}_{-0.04}$ &         $3.646^{+0.014}_{-0.015}$ \\
               & logN                                     &     $\text{cm}^{-3}$ & $19.99^{+0.01\text{(p)}}_{-0.05}$ &            $18.6^{+0.2}_{-0.1}$ &           $18.58^{+0.17}_{-0.05}$ \\
               & Afe                                      &                      &      $10.00^{\text{(p)}}_{-0.07}$ &      $10.0^{\text{(p)}}_{-0.3}$ &        $10.0^{\text{(p)}}_{-0.3}$ \\
               & norm                                     &                      &                 $0.0195\pm0.0003$ &       $0.029^{+0.001}_{-0.002}$ &      $0.0437^{+0.0007}_{-0.0008}$ \\
kerrbb         & $\spin$\tablenotemark{b}                                        &                      &   $0.99964^{+0.00007}_{-0.00003}$ & $0.9999^{\text{(p)}}_{-0.0003}$ & $0.99990^{\text{(p)}}_{-0.00004}$ \\
               & $\dot{M}$                                      & $10^{18}$~g s$^{-1}$ &                   $0.245\pm0.002$ &       $0.201^{+0.002}_{-0.001}$ &      $0.2183^{+0.0008}_{-0.0005}$ \\
               & $f_{\rm col}$                            &                      &                          1.55~(f) &                        1.55~(f) &                          1.55~(f) \\ 
xillverCP      & logN                                     &     $\text{cm}^{-3}$ &              $16.0^{+1.1}_{-0.2}$ & $15.2^{+0.8}_{-0.2 \text{(p)}}$ &           $15.98^{+0.06}_{-0.24}$ \\
               & norm                                     &                      &      $0.0032^{+0.0001}_{-0.0003}$ &    $0.0045^{+0.0009}_{-0.0005}$ &      $0.0045^{+0.0004}_{-0.0003}$ \\
NuSTAR/FPMA    & constant                                   &                      &                   $1.008\pm0.002$ &                 $1.005\pm0.002$ &                   $1.014\pm0.002$ \\
MLI            & covfrac                                  &                      &            $0.82^{+0.01}_{-0.01}$ &                   $0.81\pm0.02$ &         $0.826^{+0.010}_{-0.006}$ \\
INTEGRAL/IBIS  & constant                                   &                      &                               --- &                   $1.16\pm0.02$ &         $1.528^{+0.014}_{-0.013}$ \\
AstroSat/SXT   & constant                                   &                      &                   $0.890\pm0.006$ &                             --- &                               --- \\
               & Gain fit slope                           &                      &         $1.033^{+0.003}_{-0.002}$ &                             --- &                               --- \\
               & Gain fit offset                          &                      &         $0.011^{+0.007}_{-0.006}$ &                             --- &                               --- \\
AstroSat/LAXPC & constant                                   &                      &            $0.88^{+0.02}_{-0.03}$ &                             --- &                               --- \\
               & Gain fit slope                           &                      &                     $0.99\pm0.02$ &                             --- &                               --- \\
               & Gain fit offset                          &                      &           $-0.05^{+0.09}_{-0.07}$ &                             --- &                               --- \\
IXPE/DU1       & constant                                   &                      &                   $0.847\pm0.005$ &       $0.867^{+0.008}_{-0.010}$ &         $0.912^{+0.007}_{-0.009}$ \\
               & Gain fit slope                           &                      &                   $0.961\pm0.002$ &       $1.022^{+0.004}_{-0.002}$ &                   $1.045\pm0.002$ \\
               & Gain fit offset                          &                      &        $0.0314^{+0.008}_{-0.009}$ &      $-0.107^{+0.010}_{-0.006}$ &                  $-0.157\pm0.009$ \\
IXPE/DU2       & constant                                   &                      &         $0.815^{+0.007}_{-0.005}$ &       $0.845^{+0.007}_{-0.009}$ &                   $0.875\pm0.008$ \\
               & Gain fit slope                           &                      &       $0.9700^{+0.0015}_{-0.0016}$ &       $0.981^{+0.003}_{-0.002}$ &                   $1.032\pm0.002$ \\
               & Gain fit offset                          &                      &          $0.049^{+0.007}_{-0.008}$ &      $-0.005^{+0.010}_{-0.012}$ &                  $-0.109\pm0.009$ \\
IXPE/DU3       & constant                                   &                      &         $0.775^{+0.006}_{-0.005}$ &       $0.803^{+0.006}_{-0.008}$ &         $0.838^{+0.007}_{-0.004}$ \\
               & Gain fit slope                           &                      &                               --- &       $1.001^{+0.003}_{-0.002}$ &                   $1.031\pm0.002$ \\
               & Gain fit offset                          &                      &                               --- &      $-0.014^{+0.010}_{-0.011}$ &                  $-0.096\pm0.009$ \\
NICER          & constant                                   &                      &                         $1.0$ (f) &                             --- &                         $1.0$ (f) \\
               & Strength (1.74 keV)                      &                      &     $-0.0022^{+0.0002}_{-0.0004}$ &                             --- &                $-0.0010\pm0.0003$ \\
               & MaxTau (1.84 keV)                        &                      &         $0.039^{+0.002}_{-0.003}$ &                             --- &         $0.047^{+0.002}_{-0.001}$ \\
               & wind $\nh$\tablenotemark{c}                  &   $10^{22}$~cm$^{-2}$ &                  $1.88$ -- $4.73$ &                             --- &                  $1.48$ -- $5.51$ \\
               & $f_{\rm sc}$ \tablenotemark{c}       &                      &                $0.037$ -- $0.051$ &                             --- &                $0.076$ -- $0.139$ \\
               & Relative reflection\tablenotemark{c} &                      &                  $0.77$ -- $1.12$ &                             --- &                    $0.79$--$1.25$ \\
               & $\dot{M}$\tablenotemark{c}                 & $10^{18}$~g s$^{-1}$ &                $0.258$ -- $0.280$ &                             --- &                  $0.220$--$0.266$ \\
FIT STATISTIC  & $\chi^2$ / DOF                           &                      &                $3532.39$ / $2740$ &                $896.67$ / $783$ &                $4133.97$ / $3799$ \\
\enddata 
\tablenotetext{a}{At such high electron temperatures, the diffusion approximation built into the nthcomp model is insufficient, which may account for residuals at $\gtrsim 50$~keV in Fig.~\ref{fig:specfit}.  While not included in our model, those residuals can be well-fitted using an ad hoc broad Gaussian, with negligible impact on the other fit parameters.}
\tablenotetext{b}{The spin uncertainty is adopting a fixed $M$, $i$, and $D$ and does not include measurement uncertainty from those terms, nor any model systematic uncertainty.}
\tablenotetext{c}{Parameters are left to vary in each \NICER\ GTI; ranges shown depict the GTI ensemble.}
\tablecomments{Best fit and 90\% confidence intervals for our preferred fully-relativistic model applied to Epochs 3, 4, and 5.  Parameters with (t) are tied to others, as described in the text, while those marked (f) are fixed due to a lack of low-energy spectral coverage in Epoch 4. Any value pegged at a hard limit is marked (p).  Relative reflection describes a per GTI scale-factor for both reflection components.}  
\label{tab:results} 
\end{deluxetable} 